\documentstyle[11pt,epsf]{article}

\advance \topmargin by -\headheight
\advance \topmargin by -\headsep

\evensidemargin \oddsidemargin
\newfont{\headfont}{cmbx10 scaled 1440}
\newfont{\namefont}{cmr10}
\newfont{\initialfont}{cmr10 scaled 1200}
\newfont{\addfont}{cmti7 scaled 1440}
\newfont{\boldmathfont}{cmbx10}
\newfont{\figfont}{cmr7 scaled 1200}
\def\A{{\bf{A}}}
\def\r{{\bf{r}}}

\def\p{{\bf{p}}}
\def\ie{{\it{i.e.\space}}}
\def\H{{\cal{H}}}
\def\fun{I(\rho,\theta,\{\nu\})}
\def\epi{e^{i\pi/4}}
\def\emi{e^{-i\pi/4}}
\def\sqr2{\sqrt{2}}
\def\frnu{\{\nu\}}
\def\eir{e^{i\rho}}
\def\psitr{\Psi_{\rm{tr}}(r,\theta;\nu)}
\def\psisc{\Psi_{\rm{sc}}(r,\theta;\nu)}

\newcommand{\IZ}{{Z \kern -0.67em Z}}
\let\nopictures=Y
\newfont{\headfontb}{cmbx10 scaled 1728}
\begin{document}
\begin{titlepage}
\renewcommand{\thefootnote}{\fnsymbol{footnote}}
\begin{center}
{\headfontb Calculation of the Aharonov-Bohm\\ ~wave function}
\footnote{This work is supported in part by funds provided by the
U.~S. Department of Energy (D.O.E.) under cooperative research agreement
\#DE-FC02-94ER40818.}

\end{center}
\vskip 0.3truein
\begin{center}
{
{\Large M.}{} {\Large A}{lvarez\footnote{Email: marcos@mitlns.mit.edu}}
}
\end{center}
\begin{center}
{\addfont{Center for Theoretical Physics,}}\\
{\addfont{Massachusetts Institute of Technology}}\\
{\addfont{Cambridge, Massachusetts 02139 U.S.A.}}
\end{center}
\vskip 0.5truein
\begin{center}
\bf ABSTRACT
\end{center}
A calculation of the Aharonov-Bohm wave function is presented. The result is
a
series of confluent hypergeometric functions which is finite at the forward
direction.

\vskip 7truecm
\leftline{MIT--CTP--2473  \hfill October 1995}
\smallskip
\leftline{hep-th/9510085}

\vskip 0.5truein
\begin{center}
\bf{Submitted to Physical Review D}
\end{center}

\end{titlepage}
\setcounter{footnote}{0}

%
%

\section{Introduction}
The scattering of charged particles by an infinitely long straight solenoid
that encloses a magnetic flux, known as the Aharonov-Bohm effect, is of
paramount interest in quantum physics \cite{aha}. Even though the region
containing the magnetic field is inaccessible to the particles, the magnetic
flux inside the solenoid affects their propagation. The observed interference
pattern cannot be explained within classical physics; it is a purely quantal
effect, without classical correspondence. This fact singularizes the
Aharonov-Bohm (AB) effect from other important processes, like Rutherford
scattering.

The AB effect has been analyzed by many different approaches. We shall restri
ct
ourselves to the simplest situation of a straight filiform solenoid with
constant magnetic flux. The object of interest is the wave function of the
scattered particles and the ensuing scattering amplitude. There is
some freedom in choosing the initial wave function, but we shall consider onl
y
two possibilities: a plane wave or a well-localized wave packet. In the first
case we can drop the time dependence of the wave functions without losing
any information; this situation will be
referred to as ``time-independent scattering''. In the second case we shall
speak of ``time-dependent scattering''. Most of the early analysis of the AB
scattering can be considered time-independent.

The scattering
amplitude was first calculated by Aharonov and Bohm \cite{aha} by means of a
decomposition of the wave function in positive, negative and zero angular
momentum components, which are calculated separately. The long-distance form
of
the scattered wave far from the forward direction defines the scattering
amplitude that bears these authors' names. They did not determine the
wave function along the forward direction, where the scattering amplitude see
ms
to diverge.

A different procedure is due to Berry \cite{berry}, who obtained an exact
single-valued wave function by suming the wave function endowed with the Dira
c
phase factor over all contributions (``whirling waves'') representing differe
nt
windings round the flux line. It is also possible to obtain the same result b
y
including the geometrical phase factor (Berry's phase) instead of summing ove
r
whirling waves \cite{berry2}. In these works, which have provided a thorough
understanding of the AB effect, the wave function is presented as a series
of Bessel functions which is not further elaborated.

A more explicit calculation of the AB wave function was supplied by
Takabayashi \cite{taka}. An integral representation for the Bessel functions
enabled this author to evaluate the asymptotic limit of the wave function
far from the forward direction, thus reproducing the AB scattering
amplitude, and the leading term of the wave function near the forward
direction. Both the scattered and the transmitted wave were shown to be
discontinuos along the forward direction, in such a way that the total wave
function was continuous.

In the context of a more general investigation, Jackiw \cite{jack} found an
integral representation for the total AB wave function similar to the one
found in 2+1 dimensional gravity \cite{desjac}. His integral representation
allows for a natural decomposition of the total wave function in
``transmitted'' and ``scattered'' components. The transmitted wave was
evaluated explicitly, but the scattered wave was calculated only in the
asymptotic limit. The results agreed with those of \cite{taka}.

Another derivation of the AB scattering amplitude and of the wave function ne
ar
the forward direction has been provided by Dasni\`eres de Veigy and Ouvry
\cite{dasn}, this time by analyzing the action of the propagator on an incide
nt
plane wave. The procedure resembles those of \cite{taka} and \cite{jack}, and
yields identical results.

One of the few genuine time-dependent analyses of the AB effect has been
given by Stelitano in a recent work \cite{steli}. This author decomposes
the propagator in positive, negative and zero angular momentum components,
and calculates each
term independently in the asymptotic limit, following a procedure similar to
the original method of Aharonov and Bohm \cite{aha}. His initial wave functio
n
is a well-localized Gaussian wave packet that approaches the solenoid. The
final wave function is calculated as the convolution of the propagator with
the initial wave packet. The resulting scattering amplitude is the same as in
the time-independent analyses. The wave packet undergoes a self-interference
along the forward direction analogous to the one found in \cite{taka} and
\cite{dasn}.

It is clear that there is wide agreement about the results of these
calculations. What we present in this work is a complete evaluation of
the AB wave function based in the results of \cite{jack}. Our method is also
applied to calculate the AB propagator. Therefore, both time-dependent and
time-independent problems can be treated
in a similar way. We shall produce analytic results in terms of
confluent hypergeometric functions which are exact in any angular region,
thereby being valid at the forward direction. This method was
put forward recently in \cite{acg}, where it was used to calculate the wave
function in (2+1)-dimensional quantum scattering by a fixed spinless
source. It is based on Pauli's article on the diffraction of light by a
wedge limited by two perfectly reflecting planes \cite{pauli}.

This article is organized as follows. In Section 2 we shall review Jackiw's
solution of the Schr\"{o}dinger equation of a charged particle in presence of
 a
magnetic vortex \cite{jack}. In a time-independent analysis the scattered
wave function is given as an integral. The time-dependent propagator will
be shown to be proportional to the same integral, which is therefore the obje
ct
to calculate. That calculation is carried out in Section 3. The result is
applied to the determination of the time-independent wave function in
Section 4. The time-dependent analysis is presented in Section 5. The last
Section contains a discussion of the results and our conclusions.


\section{The wave function}
In this Section we shall review briefly the solution given in \cite{jack}
to the Schr\"odinger equation for the time-independent AB scattering problem.
If we consider that the magnetic vortex coincides with the $z$ axis, and that
the incoming charged particles approach perpendicularly the vortex, the
scattering process is esentially two-dimensional. In this situation a possibl
e
choice of vector potential is
\begin{equation}
{\A}({\r})={\Phi\over 2\pi}\nabla\theta={c\hbar\over e}\nu\nabla\theta,
\label{vectorpot}
\end{equation}
where $\Phi$ is the flux carried by the vortex, $\nu$ stands for the
``numerical flux'' defined as $\nu=e\Phi/2\pi\hbar c$ and $\theta$ is the pol
ar
angle of cylindrical coordinates. The Hamiltonian that defines the dynamics o
f
the system is
\begin{equation}
\H={1\over 2M}\left(\p-{e\over c}\A(\r)\right)^2
\label{hamiltoniano}
\end{equation}
with $\A(\r)$ given by Eq.~(\ref{vectorpot}). The quantum mechanical evolutio
n
of the particle-vortex system is determined by the time-dependent Schr\"oding
er
equation,
\begin{equation}
i\hbar{\partial\over\partial t}\Psi(r,\theta;\nu;t)=\H\,\Psi(r,\theta;\nu;t).
\label{schro}
\end{equation}
In a time-independent scattering process the Schr\"odinger equation
(\ref{schro}) reduces to an energy eigenvalue problem if time is factorized
in the usual way,
\begin{eqnarray}
\Psi(r,\theta;\nu;t)&=&e^{-itE/\hbar}\,\Psi_E(r,\theta;\nu) \nonumber\\
\H\Psi_E(r,\theta;\nu)&=&E\,\Psi_E(r,\theta;\nu).
\label{eigen}
\end{eqnarray}
Since the spectrum of the Hamiltonian $\H$ is continuous, with all energies
$E>0$, the interest resides in the eigenfunctions. These eigenfunctions are
found to be \cite{jack}
\begin{equation}
\Psi_E(r,\theta;\nu)=\sqrt{{M\over2\pi\hbar^2}}\left\{\sum_{j=-\infty}^{[\nu]
}
e^{-i\pi(\nu+j)/2}\,J_{\nu-j}(kr)\,e^{ij\theta}+\sum_{j=[\nu]+1}^{\infty}
e^{i\pi(\nu+j)/2}\,J_{j-\nu}(kr)\,e^{ij\theta}\right\}.
\label{yuki}
\end{equation}
where $k^2=2ME/\hbar^2$. The brackets [$\quad$] indicate integer part. The
sums can be performed with the help of the Schl\"afli representation for the
Bessel functions, whose contour of integration $C_s$ is depicted in Fig.~1
($\epsilon$ is a small positive value for Im$z$):
\begin{equation}
J_{\alpha}(x)=e^{i\alpha\pi/2}\int_{C_s}{dz\over 2\pi}\,e^{-ix\cos z}\,
e^{iz\alpha}.
\label{schlafli}
\end{equation}

\begin{figure}
\centerline{\hskip.4in\epsffile{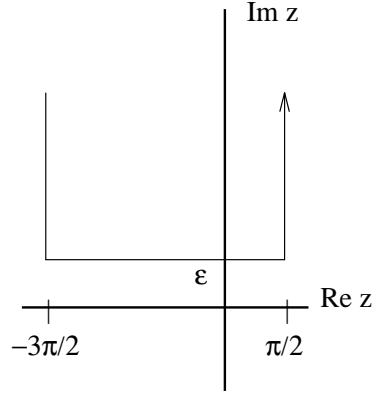}}
\caption{The Schl\"afli contour}
\end{figure}

The sums in Eq.~(\ref{yuki}) are now geometric and can be evaluated in the
usual way. After this summations and some changes of variable (we refer the
reader to \cite{jack} for details), the wave function can be represented by
\begin{equation}
\Psi_E(r,\theta;\nu)=\sqrt{{M\over2\pi\hbar^2}}\,e^{i\nu(\theta-\pi)}\int_{C}
{dz\over 2\pi}\,e^{ikr\cos(z-\theta)}\,{e^{-i\frnu z}\over 1-e^{-i z}}.
\label{contorno}
\end{equation}
The new contour $C$ is shown in Fig.~2. It is easy to verify that the
wave function (\ref{contorno}) is single-valued, \ie periodic in $\theta$ wit
h
$2\pi$ period. The contour avoids the poles $z=0\,{\rm{mod}}\,(2\pi)$. An
equivalent contour is depicted in Fig.~3. This new contour consists of a loop
that encloses the pole $z=0$, and two vertical lines. This separation
corresponds to a decomposition of the total wave function
$\Psi_E(r,\theta;\nu)$ in two components, identified as the transmitted wave
(closed contour) and the scattered wave (straight lines),

\begin{figure}
\centerline{\hskip.4in\epsffile{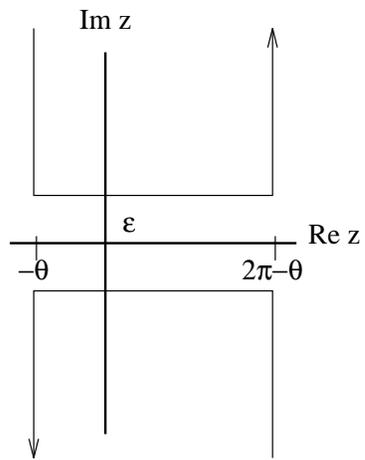}}
\caption{Contour of integration}
\end{figure}

\begin{figure}
\centerline{\hskip.4in\epsffile{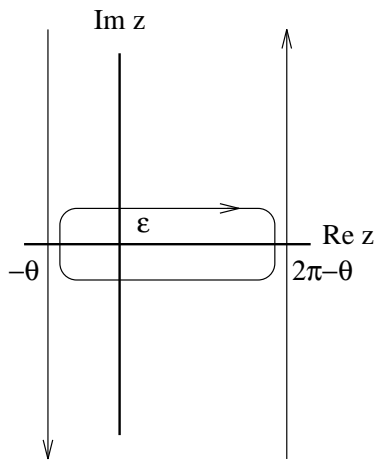}}
\caption{An equivalent contour}
\end{figure}

\begin{eqnarray}
\Psi_E(r,\theta;\nu)&=&\sqrt{{M\over2\pi\hbar^2}}\left(\psitr+\psisc\right)
\nonumber\\
\psitr&=&e^{ikr\cos\theta}e^{i\nu(\theta-\pi)} \nonumber\\
\psisc&=&e^{i[\nu]\theta}\sin(\pi\nu)\int\limits_{-\infty}^{\infty}
{dy\over\pi}\,e^{ikr\cosh y}\,{e^{\frnu y}\over e^{y-i\theta}-1}.
\label{yuki2}
\end{eqnarray}
The transmitted wave has been evaluated by Cauchy's residue theorem. The
scattered wave is an integral that can be calculated in the large $kr$ limit
by
means of a saddle point approximation {\it{except if\/}} $\theta=0$. If
$\theta=0$ the integrand develops a singularity exactly at the saddle
point $y=0$, thus invalidating the procedure.

Before proceeding to the evaluation of the integral, we demonstrate that
it also appears in the AB propagator. It has been shown in \cite{steli} that
this propagator can be written as a Bessel series:
\begin{equation}
G(r,\theta;r',\theta';\nu;t)={m\over 2\pi i\hbar t}\exp\left\{i{m\over2\hbar
t}(r^2+r'^2)\right\}\sum_{n=-\infty}^{\infty}e^{in\phi}\,e^{-i|\nu-n|\pi/2}
\,J_{|\nu-n|}\left({mrr'\over \hbar t}\right),
\label{propa}
\end{equation}
where $\phi=\theta-\theta'$. The sum can be transformed into an integral by
means of the Schl\"afli representation for the Bessel functions,
Eq.~(\ref{schlafli}). Following the same steps as for the time-independent
wave function, the AB propagator can be decomposed in two terms that
correspond to the transmitted and the scattered waves:
\begin{eqnarray}
G_{\rm{tr}}(r,\theta;r',\theta';\nu;t)&=&{m\over 2\pi i\hbar t}
\exp\left\{i{m\over2\hbar t}|\r-\r'|^2\right\}\,e^{i\nu\phi} \nonumber\\
G_{\rm{sc}}(r,\theta;r',\theta';\nu;t)&=&-{m\over 2\pi i\hbar t}
\exp\left\{i{m\over2\hbar t}(r^2+r'^2)\right\}\,e^{i[\nu]\phi}\sin(\frnu\pi)
\pi^{-1}\nonumber\\ & &\times\int\limits_{-\infty}
^{\infty}dy\,\exp\left\{i{mrr'\over\hbar t}\cosh y\right\}\,{e^{y\frnu}\over
1+e^{y-i\phi}}.
\label{green}
\end{eqnarray}
The computational difficulties lie now in the region $\phi\approx\pm\pi$, whe
re
a straightforward saddle-point approximation is ill-defined. The conclusion
of this Section is that the evaluation of both the
time-independent wave function and the time-dependent propagator reduces to
a careful calculation of the same integral, which we shall denote by $\fun$,
\begin{equation}
\fun=\int_{-\infty}^{\infty}dy\,e^{i\rho\cosh y}\,{e^{\frnu y}\over
e^{y-i\theta}-1}.
\label{muchfun}
\end{equation}


\section{Calculation of an integral}
In this Section we shall calculate the integral $\fun$ defined in
Eq.~(\ref{muchfun}). We are interested in an evaluation of this integral that
be valid even if the angular variable $\theta\in[0,2\pi)$ takes values close
to $0$ or $2\pi$.

Let us change to a new variable $s=\emi\sqr2\,\sinh{y\over 2}$, which extract
s
a Gaussian factor $\exp(-\rho s^2)$ in the integrand. The path of integration
can be taken as the real axis, so that the function $\fun$ be
\begin{eqnarray}
\fun&=&\epi\sqr2\,\eir\int_{-\infty}^{\infty} ds\,e^{-\rho
s^2}\,{1\over
1-\cos\theta +is^2}\nonumber\\ & &\times {\cos\eta-\cos\theta\over
e^{i(\eta-\theta)}-1}\,{e^{i\frnu\eta}\over\cos{\eta\over 2}}.
\label{inte}
\end{eqnarray}
We introduce now the following notation:
\begin{eqnarray}
f(s,\theta)&=&{\cos\eta-\cos\theta\over
e^{i(\eta-\theta)}-1}\,{e^{i\frnu\eta}\over\cos{\eta\over 2}}\nonumber\\
a(\theta)&=&e^{-i\pi}(1-\cos\theta).
\label{notat}
\end{eqnarray}
The choice of phase in the definition of $a(\theta)$ is not trivial since thi
s
function will appear below under a square root. The function $\fun$ is now
\begin{equation}
\fun=\emi\sqr2\,\eir\int_{-\infty}^{\infty}ds\,e^{-\rho s^2}{f(s,\theta)\over
ia(\theta)+s^2}.
\label{intdos}
\end{equation}
The obvious procedure would be to expand the whole integrand in
Eq.~(\ref{intdos}) except the Gaussian exponential, in powers of $s$
and evaluate the Gaussian integrals. The result would be ill-defined at
$\theta=0$. An alternative procedure is to expand $f(s,\theta)$ only,
according to the following formula:
\begin{equation}
f(s,\theta)=\sum_{m=0}^{\infty}e^{im{\pi\over 4}}A_m(\theta)s^m
\label{efe}
\end{equation}
This expansion is well-defined for all values of $\theta$. The odd powers of
$s$ do not contribute to the integration in Eq.~(\ref{intdos}). The functions
$A_m(\theta)$ for $m=0,2$ are
\begin{eqnarray}
A_0(\theta)&=&{1\over2}\left(e^{i\theta}-1\right) \nonumber\\
A_2(\theta)&=&{1\over8}\left(2\frnu-1\right)\left(3+e^{i\theta}-2\frnu+2\frnu
e^{i\theta}\right)
\label{as}
\end{eqnarray}
Now we insert the expansion (\ref{efe}) in Eq.~(\ref{intdos}) and consider
$\rho\neq 0$. After the change of variable $s=\tau\rho^{-1/2}$, the function
$\fun$ reads
\begin{eqnarray}
\fun&=&\emi\sqr2\,\eir\sum_{m=0}^{\infty}i^m A_{2m}(\theta)\rho^{{1\over 2}-m
}
\nonumber\\
& &\times\int_{-\infty}^{\infty}d\tau\,e^{-\tau^2}{\tau^{2m}\over
ia(\theta)\rho +\tau^2}
\label{inttres}
\end{eqnarray}
This final integration can be performed in terms of confluent hypergeometric
functions as follows:
\begin{equation}
F_2\left(1,-m+{3\over 2},ix\right)=\Gamma\left(m-{1\over2}\right)^{-1}
\int_{-\infty}^{\infty}d\tau\,e^{-\tau^2}{\tau^{2m}\over ix +\tau^2}.
\label{hyper}
\end{equation}
Therefore the final result is
\begin{eqnarray}
\fun&=&\emi\sqr2\,\eir\sum_{m=0}^{\infty}i^m A_{2m}(\theta)\rho^{{1\over 2}-m
}
\nonumber\\
& &\times\Gamma\left(m-{1\over2}\right)F_2\left(1,-m+{3\over
2},ia(\theta)\rho\right)
\label{evaluation}
\end{eqnarray}
The behaviour of the confluent hypergeometric functions defined in
Eq.~(\ref{hyper}) for large or small $|x|$ is
\begin{equation}
\begin{array}{lll}
F_2\left(1,-m+{3\over 2},ix\right)&\approx\left(m-{1\over
2}\right)(ix)^{-1}\left[1-\left(m+{1\over2}\right)(ix)^{-1}+\cdots\right]~,&
\quad |x|>\!\!>1 \nonumber\\
F_2\left(1,-m+{3\over
2},ix\right)&\approx\left[1+\left(-m+{3\over2}\right)^{-1}ix+
\left(-m+{3\over2}\right)^{-1}\left(-m+{5\over2}\right)^{-1}(ix)^2+\cdots
\right]\nonumber\\&+\pi\Gamma\left(m-{1\over2}\right)^{-1}e^{-i(m\pi/2+\pi/4)
}
x^{m-{1\over2}}e^{ix}~, &\quad |x|\approx 0~.
\end{array}
\label{limits}
\end{equation}
By means of these formulas we can evaluate $\fun$ in three limiting cases whi
ch
will be important in the next Section:
\subsection{$|\rho a(\theta)|\to\infty$}
The leading term in this limit is
\begin{equation}
\fun=-\sqrt{{\pi\over
2\rho}}\emi\eir{e^{i{\theta\over2}}\over\sin{\theta\over2}}+\cdots
\label{limit1}
\end{equation}
\subsection{$\rho>0$ and $\theta=0^+$ or $2\pi^-$}
As a shorthand we shall refer to these two possibilities as $\theta=0^+$
and $\theta=0^-$, keeping in mind that $\theta$ is an angular variable that
takes value in the interval $[0,2\pi)$. The leading contributions to $\fun$
are:
\begin{equation}
I(\rho, 0^{\pm}, \frnu)=\pm i\pi\eir +(2\frnu-1)\sqrt{\pi
i\over 2\rho}\,\eir +\cdots
\label{limit2}
\end{equation}
The value of the discontinuity $I(\rho, 0^+, \frnu)-I(\rho, 0^-, \frnu)$
could have been determined from (\ref{muchfun}) by means of a simple
distributional calculation based on the identity
\begin{equation}
{1\over 1-e^{-y}\mp i\epsilon}={\cal{P}}{1\over 1-e^{-y}}\pm\pi i\delta(y),
\label{distri}
\end{equation}
where $\epsilon\to 0^+$ and ${\cal{P}}$ denotes the principal value. The
${\cal{O}}(\rho^{-1/2})$ term in the right side
of Eq.~(\ref{limit2}) vanishes if $\frnu=1/2$. That this is exact follows
from a direct calculation using Eq.~(\ref{distri}):
\begin{equation}
I(\rho, 0^{\pm}, {1\over2})=\pm i\pi\eir
\label{half}
\end{equation}
\subsection{$\rho=0$ and $\frnu\neq 0$}
This last case can be calculated directly from the definition of $\fun$ and t
he
identity (\ref{distri}). The result is
\begin{equation}
I(0, \theta, \frnu)=-\pi e^{i\frnu\theta}e^{-i\pi\nu}{1\over \sin(\pi\nu)}.
\label{rocero}
\end{equation}
To conclude this Section we wish to remark that we have found a unified
expression for $\fun$ that embraces all particular cases as limiting values o
f
the parameters $\theta$ and $\frnu$, with the only exception of $\rho=0$. We
shall show in the next Section how these results can be succesfully applied t
o
the calculation of the AB wave function.

\section{Application to time-independent AB scattering}
We recall the solution to the static AB scattering problem found in
\cite{jack}:
\begin{eqnarray}
\psitr&=&e^{i\nu(\theta-\pi)}e^{ikr\cos\theta} \nonumber\\
\psisc&=&e^{i[\nu]\theta}\sin(\pi\nu)\pi^{-1}\fun,
\label{psis}
\end{eqnarray}
where $\fun$ is the function evaluated in the preceding Section. Therefore we
can proceed directly to calculate the scattered wave function $\psisc$ in som
e
cases of interest.

\subsection{$kr(1-\cos\theta)\to\infty$}
This is the large $kr$ limit away from the classical scattering direction
$\theta=0$. Using Eq.~(\ref{limit1}) we find
\begin{equation}
\psisc=\sqrt{i\over2\pi kr}\exp\left\{i[\nu]\theta
+i\,{\theta+\pi\over2}\right\}{\sin(\pi\nu)\over\sin{\theta\over2}}\,e^{ikr}
+\cdots
\label{asympt}
\end{equation}
The large-$r$ asymptote of $\psisc$ defines the scattering amplitude
$f(k,\theta)$ through the formula
\begin{equation}
\Psi_{\rm{sc}}(r,\theta;\nu)\stackrel{\scriptscriptstyle r\rightarrow\infty}
{\longrightarrow}\sqrt{{i\over r}}\,f(k,\theta)\,e^{ikr}.
\label{defscam}
\end{equation}
Comparing Eqs.~(\ref{asympt}) and (\ref{defscam}) we extract the usual AB
scattering amplitude:
\begin{equation}
f(k,\theta)={1\over\sqrt{2\pi k}}\exp\left\{i\left([\nu]+{1\over 2}\right)
\theta+i{\pi\over 2}\right\}{\sin(\pi\nu)\over\sin{\theta\over 2}}
\label{scattamp}
\end{equation}
Other definitions of the scattering amplitude are possible, depending on what
is meant by ``incoming'' and ``scattered'' waves. For instance we
may consider that the incoming wave function is a plane wave
$e^{ikr\cos\theta}$, and define the scattered wave as
\begin{equation}
{\tilde{\Psi}}_{\rm{sc}}(r,\theta;\nu)=\psisc+\psitr -e^{ikr\cos\theta}.
\label{alter}
\end{equation}
\ie the total wave function given by Eq.~(\ref{yuki2}), minus the incoming
plane wave. As explained in \cite{jack}, the large $r$ behaviour of this new
scattered wave defines a scattering amplitude that differs from
(\ref{scattamp}) in a $\delta(\theta)$-function, thus reproducing Ruijsenaar'
s
result in Ref.~\cite{ruij}.

\subsection{$\theta=0^{\pm}$ and $r>0$}
We are going to determine the wave function at the forward direction with the
help of Eq.~(\ref{limit2}). The scattered wave develops a finite discontinuit
y,
\begin{equation}
\Psi_{\rm{sc}}(r,0^{\pm};\nu)=\pm i\sin(\pi\nu)\,e^{ikr}
+\left(2\frnu-1\right)\sqrt{{i\over 2\pi kr}}\sin(\pi\nu)\,e^{ikr}+\cdots
\label{forward}
\end{equation}
As for the transmitted wave, we find directly from Eq.~(\ref{psis}) that it
presents a matching discontinuity:
\begin{equation}
\Psi_{\rm{tr}}(r,0^+;\nu)-\Psi_{\rm{tr}}(r,2\pi^-;\nu)=-2i\sin(\pi\nu)\,
e^{ikr}.
\label{disco}
\end{equation}
Hence the total wave function $\psitr+\psisc$ is continuous at the
forward direction as it should be. It is possible to show that not only the
discontinuities in the scattered and transmitted wave functions cancel, but
also that all the discontinuities in the derivatives of one of them cancel
against the discontinuities in the derivatives of the other. The outcome is
that the analytic expression for the AB wave function presented in
Eqs.~(\ref{psis}) and (\ref{evaluation}) is smooth even at the forward
direction, and can be used to examine the behaviour of the wave function for
any value of $\theta$.

It can be said that the decomposition of the
total wave function in ``scattered'' and ``transmited'' components does not
make sense at the forward direction (or in general at the classical scatterin
g
directions, see \cite{acg} for the situation in 2+1 dimensional gravitational
scattering). We have seen that if we allow for separately discontinuous
scattered and transmitted components of the total wave function, that
decomposition can be sustained and yields a perfectly
regular expression for the total wave function. Therefore we can calculate
its value at $\theta=0$ by averaging between $0^+$ and $2\pi^-$. The result i
s
\begin{equation}
\Psi(r,0;\nu)=\cos(\pi\nu)e^{ikr}+
(2\frnu-1)\sqrt{{i\over 2\pi kr}}\sin(\pi\nu)\,e^{ikr}+\cdots
\label{psiforward}
\end{equation}

It is interesting to note that the total time-independent wave function at
the forward direction vanishes exactly if $\frnu=1/2$
as a consequence of Eq.~(\ref{half}). That value for the fractional part
of the numerical flux causes maximal scattering outside $\theta=0$, and a tot
al
extinction of the wave function at $\theta=0$. This can be understood as a
consequence of probability conservation.

\subsection{$r=0$ and $\frnu\neq 0$}
Here we shall check that the wave function vanishes at the location of the
magnetic vortex. Going back to Eq.~(\ref{psis}) and taking into account
Eq.~(\ref{rocero}) we find that the scattered and transmitted wave functions
cancel out:
\begin{eqnarray}
\Psi_{\rm{tr}}(0,\theta,\nu)&=&e^{i\nu(\theta-\pi)}  \nonumber\\
\Psi_{\rm{sc}}(0,\theta,\nu)&=&-e^{i\nu(\theta-\pi)}
\label{errecero}
\end{eqnarray}

\section{Time-dependent scattering}
A similar analysis can be performed for the time-dependent scattering problem
{}.
The incoming wave function will be now a well-localized Gaussian wave packet
of
width $\xi$ that approaches the magnetic vortex from a long distance
$r_0>\!\!>\xi$ at $t=0$, with momentum $\bf{k}:$
\begin{equation}
\Psi_{\rm{0}}(r',\theta';0)={1\over\sqrt{2\pi}\xi}\exp\left\{ikr'\cos\theta'
-{1\over4\xi^2}(r'^2+r_0^2+2rr'\cos\theta')\right\}.
\label{incoming}
\end{equation}
In what follows we shall consider that $kr_0>\!\!>1$ and $k>\!\!>r_0/\xi^2$.
The angle $\theta'$ is strongly confined at $\theta'\approx\pi$, and similarl
y
$r'\approx r_0$. These approximations are needed to render feasible the
following calculations. Therefore we shall not be able to ascertain the
behaviour of the outgoing wave packet but in an approximate
sense.

Having said that, the outgoing wave packet will be calculated as the
convolution of the initial wave function (\ref{incoming}) with the AB
propagator,
\begin{equation}
\Psi(r,\theta;\nu;t)=\int\limits_0^\infty
r'dr'\int\limits_0^{2\pi}d\theta'\,G(r,\theta;r',\theta';\nu;t)
\Psi_{\rm{0}}(r',\theta';0).
\label{convo}
\end{equation}
We are therefore interested in an explicit evaluation of the propagator
found in Sect.~2. The ``scattered'' propagator $G_{\rm{sc}}$ can be written i
n
terms of the function $\fun$ as
\begin{equation}
G_{\rm{sc}}(r,\theta;r',\theta';\nu;t)={m\over 2\pi i\hbar t}
\exp\left\{i{m\over2\hbar t}(r^2+r'^2)\right\}\,e^{i[\nu]\phi}\sin(\frnu\pi)
\pi^{-1}\,I\left({mrr'\over\hbar t},\phi+\pi,\frnu\right),
\label{fedup}
\end{equation}
where $\phi=\theta-\theta'$. Following the same steps as in the previous
Section, we can evaluate $G_{\rm{sc}}$ in the following situations:

\subsection{$\phi\approx\pi^-$ or $-\pi^+$}

This case will be referred to as the ``forward direction''. What we mean by
this needs some clarification. If we are considering the propagator alone,
there is no doubt that $\phi=\pm\pi$ does indeed correspond to the forward
propagation of an incoming particle. However, since our initial wave
packet has a finite size, the forward direction is not well defined for the
outgoing wave packet. The difficulty can be traced back to the fact
that $\theta'$ in
Eq.~(\ref{convo}) is a variable of integration, and therefore the condition
$\phi=\pm\pi$ does not define a fixed value for the outgoing
angle $\theta$. A more realistic description makes use of
the two length scales introduced in the scattering process by the initial wav
e
packet (\ref{incoming}): $\xi$ and $r_0$. Qualitatively, the relevant values
of
$\theta'$ are
\begin{equation}
\theta'\in\left[\pi-\arctan{\xi\over r_0}, \pi+\arctan{\xi\over r_0}\right]
\label{range}
\end{equation}
and therefore the condition $\phi=\pi^-$ or $-\pi^+$ means that the
outgoing angle $\theta$ is contained in the ``forward cone''
\begin{equation}
\theta\in\left[-\arctan{\xi\over r_0}, \arctan{\xi\over r_0}\right].
\label{rangito}
\end{equation}
Aside from these considerations, the situation is very similar to the
time-independent scattering at the forward direction, and similar remarks
apply. The propagator $G_{\rm{sc}}$ splits in two parts,
\begin{eqnarray}
G_{\rm{sc}}(r,\theta;r',\theta';\nu;t)&=&G_{1}^{\pm}(r,\theta;r',\theta';
\nu;t)+G_{2}(r,\theta;r',\theta';\nu;t) \nonumber\\
G_{1}^{\pm}(r,\theta;r',\theta';\nu;t)&=&
\pm{m\over2\pi\hbar t}\exp\left\{i{m\over2\hbar t}(r+r')^2\right\}
\,\sin(\nu\pi) \nonumber\\
G_{2}(r,\theta;r',\theta';\nu;t)&=&(2\frnu-1)
\sqrt{{m\over8\pi^3 i\hbar trr'}}\exp\left\{i{m\over2\hbar t}(r+r')^2\right\}
\,\sin(\nu\pi)+\cdots
\label{propsfor}
\end{eqnarray}
The double sign $\pm$ in $G_{1}$ is $+$ if $\phi\approx-\pi^+$ and is $-$ if
$\phi\approx\pi^-$. We see that there is a finite discontinuity in $G_{1}$
analogous to the one found in the time-independent analysis. This discontinui
ty
is cancelled out by a similar discontinuity in $G_{\rm{tr}}$, which is the
propagator associated to the transmitted wave:
\begin{equation}
G_{\rm{tr}}^+(r,\theta;r',\theta';\nu;t)-G_{\rm{tr}}^-(r,\theta;r',\theta';
\nu;t)=-{m\over\pi\hbar t}\exp\left\{i{m\over2\hbar t}(r+r')^2\right\}
\,\sin(\nu\pi),
\label{discoprop}
\end{equation}
where the superindices $+$ or $-$ have the same meaning as in
Eq.~(\ref{propsfor}). Thus the total propagator is continuous at the forward
direction. Also, all discontinuities in the derivatives of $G_{\rm{sc}}$ at
the forward direction cancel with those of $G_{\rm{tr}}$. The calculation
of the total wave function at $\theta=0$ can therefore be performed by
averaging between $\theta=0^+$ and $\theta=2\pi^-$, with the following result
:
\begin{equation}
\Psi(r,0;\nu;t)=\cos(\pi\nu)\,\Psi_{\rm{free}}(r,t)+\left(2\frnu-1\right)
 \sqrt{{i\over 2\pi kr}}\sin(\pi\nu)\,\Psi_{\rm{free}}^R(r,t)+\cdots
\label{harto}
\end{equation}
The notation $\Psi_{\rm{free}}(r,t)$ stands for a free wave packet that
propagates along the forward direction, not to be confused with
$\Psi_{\rm{free}}^R(r,t)$, which propagates radially,
\begin{eqnarray}
\Psi_{\rm{free}}(r,t)&=&\int\limits_0^\infty
r'dr'\int\limits_0^{2\pi}d\theta'\exp\left\{-ikr'-{1\over4\xi^2}(r'-r_0)^2
\right\}\nonumber\\
& &\times{m\over 2\pi i\hbar t}\exp\left\{i{m\over2\hbar
t}(r+r')^2\right\}\nonumber\\
\Psi_{\rm{free}}^R(r,t)&=&\int\limits_0^\infty
dr'\exp\left\{-ikr'-{1\over4\xi^2}(r'-r_0)^2\right\}\nonumber\\
& &\times\sqrt{{m\over 2\pi i\hbar t}}\exp\left\{i{m\over2\hbar
t}(r+r')^2\right\}.
\label{freeR}
\end{eqnarray}
A comparison between Eqs.~(\ref{psiforward}) and (\ref{harto}) shows that the
behaviour of the time-dependent and time-independent wave functions is
similar at the forward direction. The leading contribution is modulated by th
e
same factor $\cos(\pi\nu)$; also, the ${\cal{O}}(r^{-1/2})$-term are identica
l.

\subsection{$\phi\neq\pm\pi\,{{\rm{mod}}}(2\pi)$ and $kr\to\infty$}
The outgoing angle $\theta$ is now outside the forward cone. In this angular
region the transmitted wave is subdominant to the scattered
wave, so that it suffices to take $G_{\rm{sc}}$ as propagator. With the help
of
Eq.~(\ref{limit1}) we find that the leading contribution to $G_{\rm{sc}}$ is
\begin{equation}
G_{\rm{sc}}(r,\theta;r',\theta';\nu;t)=-\sqrt{{m\over8\pi^3 i\hbar trr'}}\exp
\left\{i{m\over2\hbar
t}(r+r')^2\right\}\,e^{i[\nu]\phi}\sin(\frnu\pi){e^{i{\phi\over2}}\over
\cos{\phi\over2}} +\cdots
\label{gedosasy}
\end{equation}
An exact evaluation of Eq.~(\ref{convo}) with $G_{\rm{sc}}$ and
$\Psi_{\rm{0}}$ given by Eqs.~(\ref{gedosasy}) and (\ref{incoming})
respectively is not known to us. In this circumstance we can resort to the
approximation of taking $\theta'=\pi$ in $G_{\rm{sc}}$, but not in the wave
function $\Psi_{\rm{0}}$. Within this approximation the leading (large $kr$)
contribution to the scattered wave function can be easily calculated, with
the following result:
\begin{equation}
\Psi_{\rm{sc}}(r,\theta;\nu;t)=\sqrt{{i\over2\pi
kr}}\exp\left\{i\left([\nu]+{1\over2}\right)\theta\right\}{i\sin(\pi\nu)\over
\sin{\theta\over2}}\,\Psi_{\rm{free}}^R(r,t),
\label{outgoing1}
\end{equation}
where $\Psi_{\rm{free}}^R(r,t)$ is the freely propagating one-dimensional
(radial) wave packet defined in Eq.~(\ref{freeR}).

Again, the AB scattering amplitude can be derived from Eq.~(\ref{outgoing1}).
The result is of course Eq.~(\ref{scattamp}).

\section{Conclusions}
The general solution \cite{jack} to the time-independent AB wave function,
Eq.~(\ref{yuki2}), has been evaluated as a series of confluent hypergeometric
functions (\ref{evaluation}, \ref{psis}). The same procedure can be
applied to the AB propagator, wherefrom the propagation of a Gaussian wave
packet has been determined. The method is inspired by a recent work
\cite{acg} on (2+1)-dimensional gravitational scattering, which in its turn
was based on Pauli's article \cite{pauli} on diffraction of light. We have
applied that procedure to determine the AB wave function far from and along
the forward direction. Our results agree with those of previous analyses, bot
h
in the time-independent or time-dependent approach \cite{taka, dasn, steli}.

The factor $\cos(\pi\nu)$ present in Eqs.~(\ref{psiforward}) and (\ref{harto}
)
is usually interpreted as a self-interference between two ``halves'' of the
incoming wave that are split by the magnetic vortex and recombined at the
forward direction. Each half carries a phase $\exp(\pm i\pi\nu)$, whose
addition produces the $\cos(\pi\nu)$. This explanation is only heuristic, sin
ce
the scattered wave is not taken into account. We have seen that the peculiar
behaviour of the scattered wave in the forward direction, Eq.~(\ref{forward})
,
is crucial in rendering the total wave function continuous and single-valued
at $\theta=0$.

On the other hand, the separation of the wave function in ``scattered'' and a
``transmitted'' components becomes arbitrary precisely at the forward
direction. The first term in Eq.~(\ref{forward}) could as well be considered
part of the transmitted wave. Our opinion is that these descriptions of the A
B
effect rely on a semiclassical conception of the wave function, which is
supposed to be separable in parts that propagate independently. The calculati
on
presented here recovers the essential unity of the wave function after summin
g
the scattered and transmitted components, and only then we find the distincti
ve
behaviour of the AB wave function in the forward direction.

\vskip .5cm
\begin{center}
{\bf{Acknowledgements}}
\end{center}
\vskip .5cm
The author is indebted to Prof. R. Jackiw for suggesting this problem. He als
o
would like to thank D. Stelitano for helpful comments, and Prof. J. Negele
and the CTP for hospitality. This work has been partially supported by the
Spanish Ministerio de Educaci\'on y Ciencia.


\end{document}